\documentclass[a4paper,11pt]{article}
\usepackage{pos}
\usepackage{caption}
\usepackage{subcaption}
\usepackage{titlesec}

\usepackage{hyperref}
\usepackage{booktabs} 
\usepackage{makecell}

\usepackage{tikz}
\usetikzlibrary{arrows,shapes,tikzmark,decorations.pathmorphing}
\usepackage[compat=1.1.0]{tikz-feynman}

\usepackage{diagbox}
\usepackage{adjustbox}
\newcolumntype{L}[1]{>{\raggedright\let\newline\\\arraybackslash\hspace{0pt}}m{#1}}
\newcolumntype{C}[1]{>{\centering\let\newline\\\arraybackslash\hspace{0pt}}m{#1}}
\newcolumntype{R}[1]{>{\raggedleft\let\newline\\\arraybackslash\hspace{0pt}}m{#1}}
\usepackage{multirow, bigdelim}
\usepackage{graphicx} 

\usepackage{xcolor} 
\definecolor{darkred}{rgb}{0.6, 0, 0}

\tikzset{
quark/.style={postaction={decorate}, decoration={markings}},
scalar/.style={dashed,postaction={decorate}, decoration={markings,mark=at position .5 with {\arrow[#1]{latex}}}},
gluon/.style={decorate,decoration={coil,amplitude=3pt, segment length=4.7pt, pre length=.01cm, post length=.01cm}},
gluont/.style={decorate,decoration={coil,amplitude=3pt, segment length=3.50pt, pre length=.01cm, post length=.01cm}},
}

\usepackage[utf8]{inputenc}
\DeclareUnicodeCharacter{27F9}{\ensuremath{\Longrightarrow}}
\DeclareUnicodeCharacter{2209}{\ensuremath{\notin}}
\DeclareUnicodeCharacter{266D}{\ensuremath{\flat}}
\DeclareUnicodeCharacter{2208}{\ensuremath{\in}}
\DeclareUnicodeCharacter{00E9}{\'{e}}
\DeclareUnicodeCharacter{2264}{\ensuremath{\leq}}
\DeclareUnicodeCharacter{2265}{\ensuremath{\geq}}
\DeclareUnicodeCharacter{2200}{\ensuremath{\forall}}
\DeclareUnicodeCharacter{221E}{\ensuremath{\infty}}
\DeclareUnicodeCharacter{221D}{\ensuremath{\propto}}
\DeclareUnicodeCharacter{210F}{\ensuremath{\hbar}}
\DeclareUnicodeCharacter{2020}{\ensuremath{\dagger}}
\DeclareUnicodeCharacter{03B1}{\ensuremath{\alpha}}
\DeclareUnicodeCharacter{03B2}{\ensuremath{\beta}}
\DeclareUnicodeCharacter{0393}{\ensuremath{\Gamma}}
\DeclareUnicodeCharacter{03B3}{\ensuremath{\gamma}}
\DeclareUnicodeCharacter{03B8}{\ensuremath{\theta}}
\DeclareUnicodeCharacter{21D2}{\ensuremath{\Rightarrow}}
\DeclareUnicodeCharacter{0394}{\ensuremath{\Delta}}
\DeclareUnicodeCharacter{03B4}{\ensuremath{\delta}}
\DeclareUnicodeCharacter{03C1}{\ensuremath{\rho}}
\DeclareUnicodeCharacter{03BE}{\ensuremath{\xi}}
\DeclareUnicodeCharacter{03C0}{\ensuremath{\pi}}
\DeclareUnicodeCharacter{03A0}{\ensuremath{\Pi}}
\DeclareUnicodeCharacter{03BC}{\ensuremath{\mu}}
\DeclareUnicodeCharacter{03BD}{\ensuremath{\nu}}
\DeclareUnicodeCharacter{222B}{\ensuremath{\int}}
\DeclareUnicodeCharacter{2211}{\ensuremath{\Sigma}}
\DeclareUnicodeCharacter{03C3}{\ensuremath{\sigma}}
\DeclareUnicodeCharacter{03C4}{\ensuremath{\tau}}
\DeclareUnicodeCharacter{03BB}{\ensuremath{\lambda}}
\DeclareUnicodeCharacter{03B7}{\ensuremath{\eta}}
\DeclareUnicodeCharacter{03A6}{\ensuremath{\Phi}}
\DeclareUnicodeCharacter{03D5}{\ensuremath{\phi}}
\DeclareUnicodeCharacter{03A8}{\ensuremath{\Psi}}
\DeclareUnicodeCharacter{03C8}{\ensuremath{\psi}}
\DeclareUnicodeCharacter{03F5}{\ensuremath{\epsilon}}
\DeclareUnicodeCharacter{2202}{\ensuremath{\partial}}
\DeclareUnicodeCharacter{220F}{\ensuremath{\prod}}
\DeclareUnicodeCharacter{03C9}{\ensuremath{\omega}}
\DeclareUnicodeCharacter{2260}{\ensuremath{\neq}}
\DeclareUnicodeCharacter{22C5}{\ensuremath{\cdot}}
\DeclareUnicodeCharacter{27E8}{\ensuremath{\langle}}
\DeclareUnicodeCharacter{27E9}{\ensuremath{\rangle}}
\DeclareUnicodeCharacter{2192}{\ensuremath{\rightarrow}}
\DeclareUnicodeCharacter{00B2}{\ensuremath{{}^2}}
\DeclareUnicodeCharacter{221A}{\ensuremath{\sqrt{}}}
\DeclareUnicodeCharacter{223C}{\ensuremath{\sim}}
\DeclareUnicodeCharacter{226A}{\ensuremath{\ll}}
\DeclareUnicodeCharacter{00D7}{\ensuremath{\times}}

\title{Two-Loop Five-Point One-Mass Amplitudes \\ in the Spinor-Helicity Formalism}
\ShortTitle{Two-Loop Five-Point One-Mass Spinor-Helicity Amplitudes}

\author*[a]{Giuseppe De Laurentis}

\affiliation[a]{Higgs Centre for Theoretical Physics, University of Edinburgh, \\
  Edinburgh, EH9 3FD, United Kingdom
}

\emailAdd{giuseppe.delaurentis@ed.ac.uk}

\abstract{
  Processes involving electroweak vector bosons in association with
  jets are crucial for precision studies of the Standard Model at the
  Large Hadron Collider. Accurate predictions for the process
  $pp\rightarrow V(\rightarrow\bar\ell\ell)jj$ at
  next-to-next-to-leading order (NNLO), as well as N$^3$LO ones for
  the process with one fewer jet, require two-loop five-point one-mass
  amplitudes. These exhibit a significant increase in complexity
  compared to their massless counterparts. We present results for the
  partonic processes $\bar qQ\bar Qq\bar\ell\ell$ and $\bar
  qggq\bar\ell\ell$, with a forthcoming package for phenomenology. The
  present recalculation is based on the previous analytic results from
  ref.~\cite{Abreu:2021asb}. For the first time, we express two-loop
  five-point one-mass amplitudes in the spinor-helicity formalism. We
  achieve three orders of magnitude smaller expressions compared to
  previous results. This will facilitate phenomenological studies and
  help clarify mathematical properties of the amplitude.
  }

\FullConference{%
  42$^\text{nd}$ International Conference on High Energy Physics - ICHEP2024,\\
  17-24 July, 2024\\
  Prague, Czech Republic
}

\begin{document}
\maketitle

\section{Introduction}
\vspace{-2mm}
Processes involving electroweak massive vector bosons in association
with jets play a key role for precision studies at the Large Hadron
Collider (LHC). The $W$ and $Z$ boson not only serve as standard
candles, but they are also of great interest for the study of the
electroweak sector, e.g.~in relation to the electroweak phase
transition or the matter-antimatter asymmetry problem.

Precision studies require next-to-next-to-leading-order (NNLO) cross
sections, which in turn necessitate two-loop five-point one-mass
amplitudes. These are the current state of the art for
high-multiplicity multi-loop amplitudes and are available analytically
only in the leading-color approximation. In fact, integration-by-parts
reduction onto a basis of master integrals and subsequent analytic
reconstruction of their coefficients remain challenging problems, even
if a complete basis of transcendental functions is known, thanks to
five-point one-mass pentagon functions \cite{Chicherin:2021dyp,
  Abreu:2023rco}.

Analytic two-loop amplitudes are known for $pp \rightarrow Wb\bar{b}$
\cite{Badger:2021nhg,Hartanto:2022qhh}, $pp \rightarrow Hb\bar{b}$
\cite{Badger:2021ega}, $pp \rightarrow Wjj$ \cite{Abreu:2021asb}, $pp
\rightarrow Wj\gamma$ \cite{Badger:2022ncb}, and, most recently, $pp
\rightarrow W\gamma\gamma$ \cite{Badger:2024sqv}. The latter for the
first time also includes numerical evaluations of the subleading-color
amplitudes. Some of these are also valid for a $Z$ or $\gamma^*$ boson
instead of a $W$ boson, with the omission of gauge-invariant
non-planar contributions to the leading-color amplitude. Out of all of
these processes, NNLO cross sections have only been computed for $pp
\rightarrow Wb\bar{b}$ \cite{Hartanto:2022qhh, Hartanto:2022ypo},
including leading $b$-mass effects \cite{Buonocore:2022pqq}, and $pp
\rightarrow Zb\bar{b}$ \cite{Mazzitelli:2024ura}. This is at least
partially due to the complexity of the expressions for the
double-virtual contributions.

In an upcoming publication \cite{DeLaurentis:2024xxx}, we re-express
the two-loop planar amplitudes for $pp \rightarrow Vjj$ from
Mandelstam to spinor-helicity variables, simplifying the rational
coefficients from $\sim$1.4 GB to $\sim$1.9 MB. Other key differences
are direct reconstruction of the six-point helicity amplitudes, rather
than form factors, and a basis change in the space of
pentagon-function coefficients \cite{DeLaurentis:2024xxx,
  DeLaurentis:2023nss}. A first reconstruction is performed with
spinor-ansatz methods \cite{DeLaurentis:2019bjh} via finite-field
samples \cite{vonManteuffel:2014ixa, Peraro:2016wsq}. A second
reconstruction to obtain simpler results is performed via
$p\kern0.1mm$-adic evaluations \cite{DeLaurentis:2022otd,
  Chawdhry:2023yyx}.
\vspace{-1mm}

\section{Helicity Amplitudes and Finite Remainders}
\begin{figure}[t]
\centering
\hspace{0.02\textwidth}
\begin{minipage}{0.38\textwidth}
    \centering
    \includegraphics[width=\linewidth]{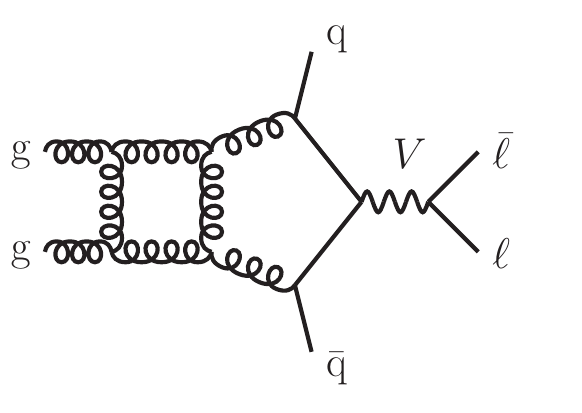}
\end{minipage}\hspace{0.08\textwidth}
\begin{minipage}{0.38\textwidth}
    \centering
    \includegraphics[width=\linewidth]{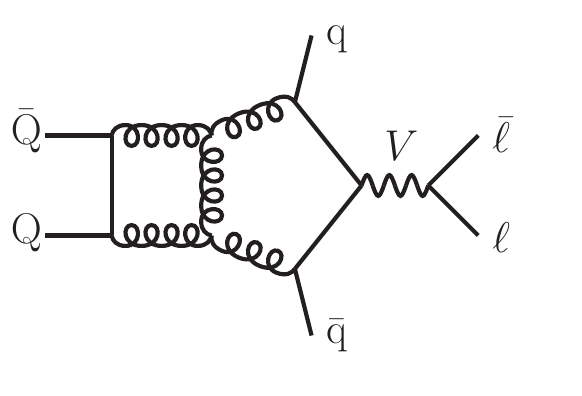}
\end{minipage}
\vspace{-3mm}
\caption{Representative two-loop diagrams for the two partonic
  processes contributing to $pp\rightarrow Vjj$.}
\label{fig:diagrams}
\vspace{-2mm}
\end{figure}

\vspace{-2mm}
We compute perturbative corrections to the process
$pp\rightarrow V(\rightarrow\ell\bar\ell)jj$, to which two distinct
partonic channels contribute,
\vspace{-1mm}
\begin{gather}
  0 \rightarrow \bar q_{p_1}^{h_1} \, Q_{p_2}^{h_2} \, \bar Q_{p_3}^{h_3} \, q_{p_4}^{h_4} \, V_{p_V} (\rightarrow  \bar\ell_{p_5}^{h_5} \, \ell_{p_6}^{h_6}) \, , \label{eq:4q1W} \\
  0 \rightarrow \bar q_{p_1}^{h_1} \, g_{p_2}^{h_2} \, g_{p_3}^{h_3} \, q_{p_4}^{h_4} \, V_{p_V} (\rightarrow \bar\ell_{p_5}^{h_5} \, \ell_{p_6}^{h_6})  \, .\label{eq:2q2g1W}
\end{gather}
We work in the all-outgoing convention, as indicated by momentum
($p_i$) and helicity ($h_i$) labels. We compute corrections up to
order $O(\alpha_s^2)$, with representative diagrams shown in figure
\ref{fig:diagrams}. Two-loop corrections are considered in the planar
approximation, which corresponds to the leading-color $Wjj$\linebreak amplitude,
while it omits gauge-invariant non-planar contributions to the
leading-color $Zjj$ one.

There are two independent helicity assignments for the four-quark
amplitude, both are next-to-maximally-helicity-violating (NMHV),
\begin{gather}
  0 \rightarrow \bar q_{p_1}^{+} \, Q_{p_2}^{+} \, \bar Q_{p_3}^{-} \, q_{p_4}^{-} \, \bar\ell_{p_5}^{+} \, \ell_{p_6}^{+} \quad \text{and} \quad  0 \rightarrow \bar q_{p_1}^{+} \, Q_{p_2}^{-} \, \bar Q_{p_3}^{+} \, q_{p_4}^{-} \, \bar\ell_{p_5}^{+} \, \ell_{p_6}^{+} \, , \label{eq:q-hels}
\end{gather}
while the two-quark two-gluon amplitude has three, the first is MHV,
the latter two are NMHV,
\begin{gather}
  0 \rightarrow \bar q_{p_1}^{+} \, g_{p_2}^{+} \, g_{p_3}^{+} \, q_{p_4}^{-} \, \bar\ell_{p_5}^{+} \, \ell_{p_6}^{-} \, , \\[1mm]
  0 \rightarrow \bar q_{p_1}^{+} \, g_{p_2}^{+} \, g_{p_3}^{-} \, q_{p_4}^{-} \, \bar\ell_{p_5}^{+} \, \ell_{p_6}^{-} \; , \; \text{and} \quad 0 \rightarrow \bar q_{p_1}^{+} \, g_{p_2}^{-} \, g_{p_3}^{+} \, q_{p_4}^{-} \, \bar\ell_{p_5}^{+} \, \ell_{p_6}^{-} \, .
\end{gather}

\pagebreak

\vspace{-1mm}
\paragraph{Finite remainder}
We construct finite remainders, $\mathcal{R}$, from renormalized
amplitudes, $\mathcal{A}_R$, by subtracting infrared singularities as
predicted by universal operators, $\boldsymbol I$, multiplied by
lower-loop amplitudes \cite{Catani:1998bh, Becher:2009cu,
  Gardi:2009qi},
\begin{equation}\label{eqn:remainder}
  {\mathcal R}^{\ell\text{-loop}} = \mathcal{A}_R^{\ell-\text{loop}} - \sum_{k=1}^{\ell} {\boldsymbol I}^{k\text{-loop}} \mathcal{A}_R^{(\ell-k)\text{-loop}} ~+~ \mathcal{O}({\epsilon}) \,,
\end{equation}
where $\epsilon$ is the dimensional regulator, $D=4-2\epsilon$. The
finite remainder can be written as a weighted sum of pentagon
functions $h_i$ \cite{Chicherin:2021dyp,
  Abreu:2023rco} with rational function coefficients $r_i$,
\begin{equation}
  \mathcal{R}(\lambda, \tilde\lambda) = \sum_i r_{i}(\lambda,\tilde\lambda) \, h_i(\lambda\tilde\lambda) \, .
\end{equation}
The arguments show that the process-dependent rational coefficient
$r_{i}$ are little-group dependent functions of the rank-one spinors,
$\lambda$ and $\tilde\lambda$, while the process-independent
transcendental functions are little-group independent functions of the
two-by-two spinors, $\lambda\tilde\lambda$, or, equivalently,
four-momenta.

We refer to the remainders for the process of eq.~\ref{eq:4q1W} as
$R_q$ and those for that of eq.~\ref{eq:2q2g1W} as $R_g$.

\section{Analytic Reconstruction over Finite Fields}
We perform a first reconstruction of the coefficients using
finite-field samples only. We set up finite-field evaluations of the
helicity remainders in \texttt{C++} from the form factor decomposition
of ref.~\cite{Abreu:2021asb}. This results in faster evaluations
compared to rerunning \texttt{Caravel} \cite{Abreu:2020xvt}, but the
compilation of the original $\sim$1.4 GB expressions is a non-trivial
task resulting in almost 20 GB of \texttt{C++} binaries. Since each
helicity amplitude was decomposed in terms of 3 form factors, moving
from the latter to the former already gives a factor 3 improvement.

\paragraph{Reduced six-point phase space}
We consider the coefficient $r_i(\lambda, \tilde\lambda)$ as functions
of six massless legs, factoring out the $M^2_V$ dependence as
$s_{56}/(s_{56}-M^2_V+i\Gamma_VM_V)$. Since the spinors associated
with the lepton pair can only enter through the decay current of the
massive vector boson $V$, namely as $[5| \sigma^\mu | 6\rangle$, we
  restrict the polynomial ring \cite{DeLaurentis:2022otd} to be,
\begin{equation}
  R_6^{\text{red.}} = \mathbb{F}\big[|1⟩, [1|, \dots, |4⟩, [4|, [5|, |6⟩ \big] \, ,
\end{equation}
where crucially $|5\rangle$ and $[6|$ are absent. This allows us to
  drop the momentum conservation equivalence relation in the ansatz
  construction by replacing every occurrence of $p_V$ for
  $\sum_{i=1}^4 p_i$.

\paragraph{Spinor alphabet}
The rational functions belong to the field of fractions of
$R_6^{\text{red.}}$. They read,
\begin{equation}
  r_i(\lambda, \tilde\lambda) = \frac{\mathcal{N}_i(\lambda, \tilde\lambda)}{\prod_j
    \mathcal{D}_j^{\alpha_{ij}}(\lambda, \tilde\lambda)} \, , \label{eq:lcd}
\end{equation}
where $\mathcal{N}_i(\lambda, \tilde\lambda)$ is a polynomial, the
$\mathcal{D}_j(\lambda, \tilde\lambda)$ are irreducible, and
$\alpha_{ij}\in \mathbb{Z}$ are minimised.

The poles $\mathcal{D}_j$ are expected to be related to the letter of
symbol alphabet \cite{Abreu:2018zmy}. We observe 44 distinct factors,
for which a representative set is
\begin{equation}
\begin{gathered}
 \{\mathcal{D}_{\{1,\dots,44\}}\} = \bigcup_{\sigma \; \subset \;
   \text{Aut}(R_6)} \sigma \circ \big\{ \langle 12 \rangle, \langle
 1|2+3|1], \langle
1|2+3|4], s_{123}, \\[-5mm]
  \kern70mm \Delta_{12|34|56}, ⟨2|3|p_V|1|2]-⟨3|4|p_V|1|3])\big\} \, ,
\end{gathered}
\end{equation}
where $\Delta_{K_1|K_2|K_3}=(K_1 \cdot K_2)^2-K_1^2K_2^2$, and the
only new denominator factor at two loops is the last one. While not
all permutations of these letters appear as denominator factors, a
significantly larger set appears as numerator zeros. These include
spinor chains involving the lepton spinors, e.g.~$⟨6|2+3|1]$,
differences of Mandelstams, e.g.~$(s_{14}-s_{23})$ or
$(s_{123}-s_{124})$, and more.

We reconstruct all least common denominators (LCD) through a pair of
univariate slices, sort the $r_i$ by polynomial degree of
$\mathcal{N}_i$, and pick the simplest $\mathbb{Q}$-independent subset
($\mathcal{B}$). Column 2 of table~\ref{tab:ansatze-sizes} shows the
ansatz size in LCD form at this stage for two of the remainders in
$\mathcal{R}_q$ and $\mathcal{R}_g$. The largest ansatz exceeds 32 M
free parameters, clearly too complex to handle in this form.

\begin{table*}[t]
\renewcommand{\arraystretch}{0.9}
\centering
\begin{tabular}{c @{\hskip 5mm} c @{\hskip 5mm} c}
\toprule
Helicity & \multicolumn{2}{c}{LCD Ansatz parameters and size}  \\
remainder & before basis change & after basis change \\
\midrule
\(\mathcal{R}_q(1_{\bar q}^{+}, 2^{-}_Q, 3^{+}_{\bar Q} , 4^{-}_q; 5^{+}_{\bar\ell}, 6^{-}_{\ell})\) & \makecell{$[86], \,\{3, -7, 7, -3, -1, 1\}$ \\ Size: $9,\kern-0.2mm 253,\kern-0.2mm 408$} & \makecell{$[33], \{1, -2, 1, -2, -1, 1\} $ \\ Size: $110,\kern-0.2mm994$} \\
\(\mathcal{R}_g(1_{\bar q}^{+}, 2^{-}_g, 3^{+}_g, 4^{-}_q; 5^{+}_{\bar\ell}, 6^{-}_{\ell})\) & \makecell{$[114], \,\{3, -12, 12, -3, -1, 1\}$ \\ Size: $32,\kern-0.2mm 602,\kern-0.2mm 941$} & \makecell{$[54], \{1, -7, 7, -1, -1, 1 \} $ \\ Size: $795,\kern-0.2mm 718$} \\
\bottomrule
\end{tabular}
\caption{
\label{tab:ansatze-sizes}
For the helicity remainders in the first column, this table shows the
LCD ansatz size of the most complicated function in the basis, before
and after basis change. Square brackets denote mass dimension, curly
brackets denote little-group weights. The ans\"atze are generated with
\texttt{lips}, \texttt{syngular}, \texttt{Singular} \cite{DGPS}, and
\texttt{OR-Tools} \cite{ortools}.  Complete data for the remaining
partials will be provided in ref.~\cite{DeLaurentis:2024xxx}.
}
\vspace{-2.5mm}
\end{table*}

\paragraph{Basis change}
We perform a basis change in the vector space $\text{span}(r_i)$, that
is, we write,
\begin{equation}
  \mathcal{R} = r_i h_i = r_{i \in \mathcal{B}} M_{ij} h_j =
  \tilde{r}_i O^{-1}_{ij} M_{jk} h_k = \tilde{r}_i \tilde{M}_{ij} h_j \, .
\end{equation}
The matrix $O_{ij}\in \mathbb{Q}$ is constructed by searching the
space of intersections of null spaces for a close to optimal choice of
poles $\mathcal{D}_k$ and orders $m$ to eliminate, while maintaining
$\text{rank}(O_{ij})=|\mathcal{B}|$,
\begin{equation}
  O_{ij} = \bigcap_{k, m} \, \text{null}_{ij}(\text{Res}(r_{i' \in
    \mathcal{B}}, \mathcal{D}_k^m)) \, .
\end{equation}
The residues are computed through formal Laurent series obtained via
additional univariate slices in $\mathbb{F}_p$. We perform the basis
change one helicity configuration at a time. Column 3 of
fig.~\ref{tab:ansatze-sizes} shows the ansatz size in LCD form after
the basis change. The functions are now clearly much simpler.

\paragraph{Partial fraction decomposition}
Even after basis change, the LCD form is too complex to
reconstruct. Since the rational coefficients are well known to be
sparse, we perform a multivariate partial fraction decomposition. We
impose constrains observed in one-loop six-point amplitudes and in
simple two-loop five-point one-mass ones. For instance, no fraction in
the partial fraction decomposition is allowed to have more than a
single three-particle Mandelstam ($s_{ijk}$), or more than a single
three-mass Gram determinant ($\Delta_{ij|kl|mn}$), or to mix $\langle
a | b + c | a ]$ with $\Delta_{ij|kl|mn}$, etc. At this stage, we do
  not verify these statements through evaluations on codimension-two
  surfaces.

We impose constraints until the ansatz sizes in partial fraction form
are around 40k. We can solve such systems in around 13 minutes by
Gaussian elimination over 32-bit $\mathbb{F}_p$ on the available RTX
2080 Ti's. The hard size limit imposed by the available VRAM (11 GB)
on this GPU is 52k.

\vspace{-1.7mm}
\section{Analytic Reconstruction with $\boldsymbol p\kern-0.2mm$-adic Numbers}
\vspace{-2mm}

At this stage we gain access to fast and stable arbitrary-precision
$p\kern0.1mm$-adic evaluations, which we perform in Python using
\href{https://github.com/GDeLaurentis/lips}{\texttt{lips}} and
\href{https://github.com/GDeLaurentis/pyadic}{\texttt{pyadic}}
\cite{DeLaurentis:2023qhd}. We also decide to recombine the pairs of
NMHV helicity configurations into a single vector space via a
$2\leftrightarrow 3$ swap. This is justified by the overlap of the
vector spaces. For instance, the two $\mathcal{R}_q$ vector spaces for
the helicities of eq.~\ref{eq:q-hels} have sizes 70 and 98
respectively, while their sum has size 142, as shown in column 3 of
table \ref{tab:VS-sizes}.

\vspace{-1.5mm}
\paragraph{Codimension-two study}
For each rational function, we study its behaviour on codimension-two
surfaces. That is, we evaluate it at points $(\eta, \tilde\eta)$ near
said surfaces and record its valuation $\nu_p$,
\vspace{-0mm}
\begin{equation}
  \nu_p(r_i(\eta, \tilde\eta)) : (\eta, \tilde\eta) \; \text{close
    to} \; V\big(\big\langle \mathcal{D}_j, \mathcal{D}_k
  \big\rangle\big) \, .
\end{equation}
\vspace{-5.5mm}

\noindent We study both surfaces associated to ideals generated by
pairs of denominator factors, to identify better partial fraction
decompositions, and ideals generated by a denominator factor and
another irreducible spinor product, to identify numerator
insertions. Both depend on whether and how quickly the numerator
$\mathcal{N}_i$ vanishes, which can be inferred from $\nu_p(r_i)$ and
$\nu_p(\prod_j D_i^{\alpha_{ij}})$.

\vspace{-1.5mm}
\paragraph{Iterated residue reconstruction}
We then proceed to iteratively reconstruct multivariate residues of
the rational coefficients, as proposed in ref.~\cite[section
  3.3]{DeLaurentis:2019bjh}. We use $p\kern0.1mm$-adic evaluations
near codimension-one and codimension-two surfaces to numerically
isolate the residues. If a reconstructed numerator is not fully
factored as a product of simple irreducible spinor brackets, then we
iterate the reconstruction on the specific fraction. This often
reveals simpler representations by identifying additional numerator
factors. This reconstruction is performed on a laptop.

\vspace{-2mm}
\section{Conclusions}
\vspace{-2mm}

\begin{figure}[h!]
\centering
\begin{minipage}{0.90\textwidth}
    \centering
    \includegraphics[width=\linewidth]{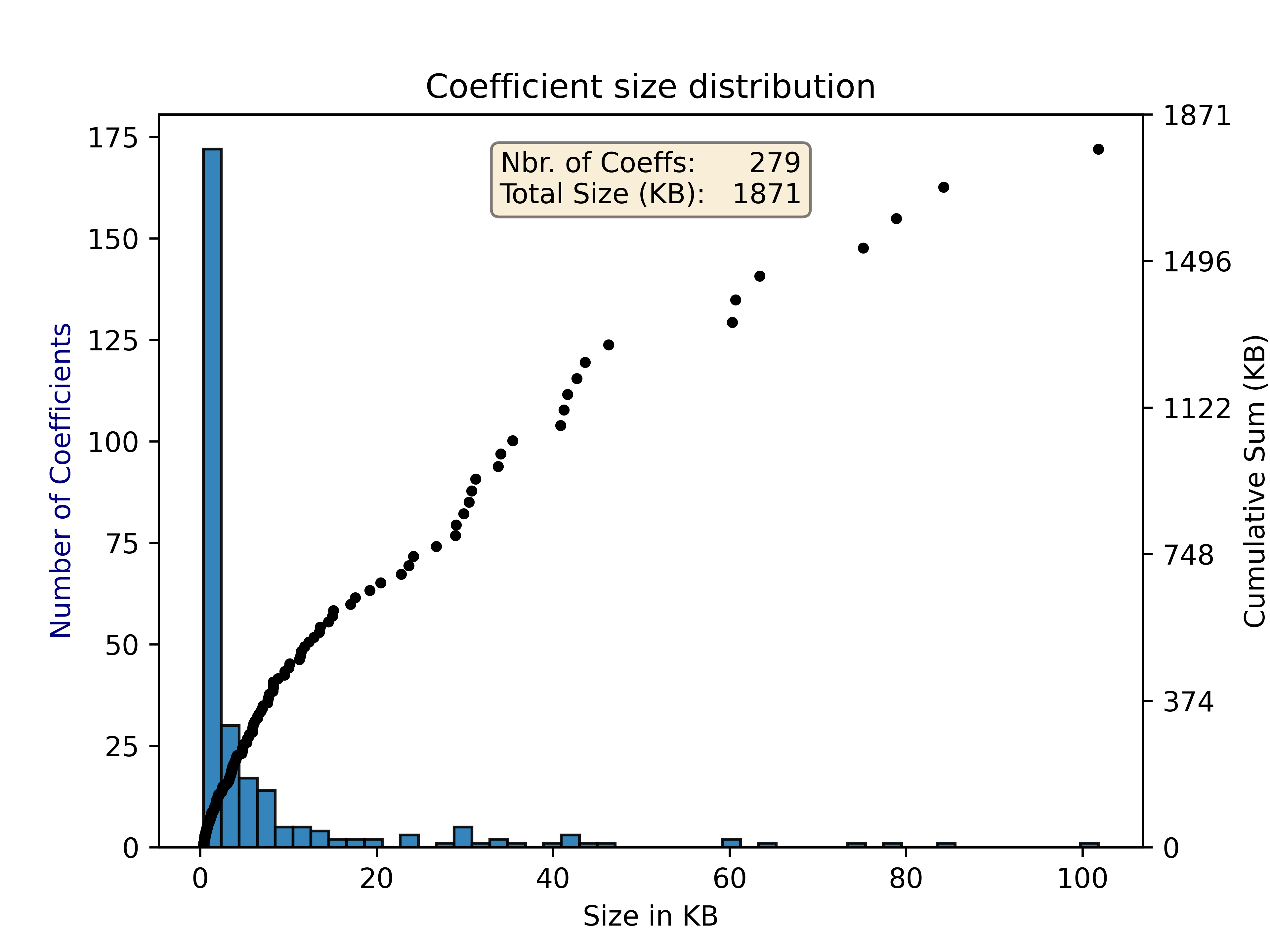}
\end{minipage}
\vspace{-3mm}
\caption{ \label{fig:size-distribution} In blue, the histogram shows the distribution in size (KB)
  of the coefficients in the bases $\tilde r$. In black, the scatter
  plot show the cumulative distribution of the size of the
  coefficients. The complexity of the results is driven by a limited
  number of complex functions (towards the right), while the majority
  of them are fairly simple (towards the left). }
\vspace{-2mm}
\end{figure}

We have obtained the first two-loop five-point one-mass amplitudes in
the spinor-helicity formalism, namely the planar corrections to
$pp\rightarrow V(\rightarrow \bar\ell\ell)jj$. These two-loop
amplitudes are now expressed in a form analogous to the corresponding
one-loop amplitudes obtained in 1997 \cite{Bern:1997sc}. \linebreak
\indent Starting from the form-factor decomposition in terms of
Mandelstam invariants of ref.~\cite{Abreu:2021asb}, we built helicity
amplitudes in a reduced six-point phase space, and reconstructed them
first through finite-field samples, and then further simplified them
via $p\kern0.1mm$-adic evaluations. Table~\ref{tab:VS-sizes} and
figure~\ref{fig:size-distribution} summarise our results for the
partonic amplitudes. Compared to the original $\sim$1.4 GB of
expressions, these results constitute a close to three order of
magnitude improvement in size. Further details and a public code for
phenomenology will be presented in an upcoming publication
\cite{DeLaurentis:2024xxx}. \linebreak \pagebreak

\paragraph{Acknowledgements}
I would like to thank H.~Ita, B.~Page, and V.~Sotnikov for comments on
this manuscript and collaboration in the related computation and
upcoming publication.

\begin{table}[h!]
\renewcommand{\arraystretch}{1.2} \centering
\begin{tabular}{C{14ex}C{10ex}C{12ex}C{10ex}C{10ex}C{10ex}}
\toprule
Process & Helicity & Vector-space dimension & Generating set size & File size ($\tilde r_i$) & Matrices Size ($\tilde M_{ij}$) \\
\midrule
$\bar qQ\bar QqV(\rightarrow \bar\ell\ell)$ & NMHV & 142 & 86 & 120 KB & 526 KB \\
$\bar qggqV(\rightarrow \bar\ell\ell)$ & MHV & 54 & 54 & 167 KB & 33 KB \\
$\bar qggqV(\rightarrow \bar\ell\ell)$ & NMHV  & 219 & 139 & $\kern-2.5mm$1$\kern0.5mm$590 KB & 953 KB \\
\bottomrule
\end{tabular}
\caption{
  \label{tab:VS-sizes} For each partonic process contributing to
  $pp\rightarrow Vjj$, and for each helicity configuration, this table
  shows the dimension of the vector space of rational functions, the
  number of functions in the generating set that spans the space upon
  closure under the symmetries of the vector of little-group scalings,
  and the size of the plain text file where the generating set is
  stored. The last column shows the size of the matrices of rational
  numbers needed to express the pentagon-function coefficients in
  terms of the basis of the vector space. }
\end{table}
\vspace{-10mm}

\pagebreak

\setlength{\bibsep}{3pt}
\bibliography{main}

\end{document}